\documentclass[aps,pre,twocolumn,superscriptaddress]{revtex4-1}

\usepackage{amsmath}
\usepackage{comment}
\usepackage{graphicx}

\begin{document}


\title{Detecting signals of weakly first-order phase transitions \\ in two-dimensional Potts models}


\author{Shumpei Iino}
\email[E-mail address: ]{iino@issp.u-tokyo.ac.jp}
\affiliation{Institute for Solid State Physics, The University of Tokyo, Kashiwa, Chiba, Japan}
\author{Satoshi Morita}
\affiliation{Institute for Solid State Physics, The University of Tokyo, Kashiwa, Chiba, Japan}
\author{Anders W. Sandvik}
\affiliation{Department of Physics, Boston University, 590 Commonwealth Avenue, Boston, Massachusetts 02215, USA}
\author{Naoki Kawashima}
\affiliation{Institute for Solid State Physics, The University of Tokyo, Kashiwa, Chiba, Japan}


\date{\today}

\begin{abstract}
  We investigate the first-order phase transitions of the $q$-state Potts models with $q = 5, 6, 7$, and $8$ on the two-dimensional square lattice, using Monte Carlo simulations. At the very weakly first-order transition of the $q=5$ system, the standard data-collapse procedure for the order parameter, carried out with results for a broad range of system sizes, works deceptively well and produces non-trivial critical exponents different from the trivial values expected for first-order transitions. However, we show a more systematic study on the `pseudo-critical' exponents as a function of the system size signals first-order phase transitions. We also derive a novel scaling behavior of Binder ratio based on a phenomenological theory for first-order transitions, which can detect the weakly first-order transitions in much smaller lattices than the correlation lengths. The results overall show that proper care is indispensable to diagnose the nature of a phase transition with limited system sizes.
\end{abstract}


\maketitle



\section{INTRODUCTION\label{sec:intro}}

While the distinction between first-order and continuous phase transitions is conceptually clear, in practice one often encounters cases where it is technically difficult to tell one from the other. At some first-order phase transitions, the correlation length at the transition point is large, and expected jumps in the energy or the entropy are hardly detectable~\cite{Binder1981_statphys}. Such phase transitions are referred to as weakly first-order ones. A classic example is the $q$-state Potts models with $q=5$ in two dimensions and $q=3$ in three dimensions~\cite{Wu1982}. In the former case the correlation length is about 2500 lattice spacings~\cite{Klumper1989,Buffenoir1993} and no discontinuities can be observed in lattices of typical size in Monte Carlo (MC) simulations.

From the viewpoint of Landau's phenomenological theory of phase transitions, the order of the transition contains important information on the structure of the phase diagram. In some cases, the distinction between the first-order and the continuous transitions may have further implications. The existence of a pseudo-transition in the metastable region was discussed for two-dimensional Potts models with $q=7$ and $q=10$~\cite{Fernandez1992}. Implications on the dynamics was discussed in the context of the electroweak phase transition of the early universe~\cite{Gleiser1994}. Furthermore, the nature of the quantum phase transition between an antiferromagnetic and a spontaneously dimerized ground state in two-dimensional quantum magnets have been actively discussed~\cite{Sandvik2007,Harada2013,Shao2016}. If this transition is continuous, it falls outside the Landau-Ginzburg-Wilson paradigm of the phase transition, since the relevant symmetry group at one side of the transition is not a subgroup of that on the other side. It is argued that such a transition may be described by the deconfined quantum-criticality phenomenon~\cite{Senthil2004}, but it has been difficult to rule out more conventional weakly first-order transitions.

The MC method is often used in studying phase transitions since it allows access to relatively large systems compared to other numerical methods. Reaching large sizes is particularly important in diagnosing weakly first-order phase transitions since, ideally, one should study systems as large as or larger than the correlation length to unambiguously determine that a transition is of first order. Thus, if the correlation length is large, it may in practice be very difficult to distinguish between the first-order and continuous cases (we note that the tensor network methods can make this possible, and actually a recent computation succeeds in differentiating weakly first-order transitions from continuous ones~\cite{Morita2018}). To analyze data from MC simulations one conventionally uses the finite-size-scaling (FSS) approach~\cite{Fisher1972}. One manifestation of FSS is that data for the order parameter, or some other singular quantity, plotted versus the control parameter (e.g., the temperature) for different system sizes fall on a single curve (the corresponding scaling function) when the systems are sufficiently large and the variables are properly scaled with the system size raised to the appropriate power associated with the relevant universality class. Therefore, procedurally one adjusts the exponents so as to obtain the best collapse of the available data points onto a single curve, within the statistical errors. When applying this procedure to first-order transition, it is generally anticipated that such collapse would be obtained with trivial, generic values known for the first-order transitions, e.g., the correlation length exponent $\nu=1/d$, $d$ being the system's dimensionality~\cite{Nienhuis1975,Fisher1982,Binder1984}. However, since this is true only when the system size exceeds the correlation length, for a weakly first-order transition one may not expect that the FSS behavior is the case in the system size less than the finite but large correlation length. Moreover, it has been proposed that `pseudo-critical' scaling with non-trivial exponents may be observed for smaller system sizes than the correlation length~\cite{Wang2017}, and, therefore, wrong conclusions may be drawn based on studies where sufficiently large systems have not been reached.

In this article, we demonstrate that pseudo-critical behavior can indeed be observed when standard data-collapse procedures are employed with data for very weakly first-order transitions. We focus on the two-dimensional $q$-state Potts model, with $q=5,6,7,8$, as a model whose phase transition is weakly first-order for $q=5$ and becomes more strongly discontinuous as $q$ is increased. For $q=5$, good scaling collapse of the near-critical order parameter can be achieved over a wide range of system sizes, with exponents that are quite far from their expected first-order values. However, we cannot clearly establish the connection between the pseudo-scaling behavior observed in the present article and the influence of fictitious fixed points discussed in Ref.~\onlinecite{Wang2017}. We will give some discussion in the last part of the present article. Eventually the trivial first-order values are approached exponentially rapidly as the system size grows, which we can observe clearly for $q>6$.

The outline of the rest of the paper is as follows: In Sec.~\ref{sec:model} we explain the methods of computation and physical quantities calculated; the order parameter and the associated Binder ratio. We show two ways of detecting the sign of weakly first-order transitions even in the smaller system sizes than the correlation lengths in Sec.~\ref{sec:exponent} and Sec.~\ref{sec:binder}. In Sec.~\ref{sec:exponent} we first present the deceptively successful results of data collapse analysis. However, we demonstrate studying the size-dependent effective exponents makes it possible to detect a sign of first-order transition. In Sec.~\ref{sec:binder} a novel characteristic of the Binder ratio for a first-order phase transition is derived based on a phenomenological theory, which also enables us to differentiate weakly first-order transitions from continuous ones. In Sec.~\ref{sec:conclusions} we briefly summarize and further discuss their significance and implications.
%

\section{MODEL AND SIMULATION METHODS\label{sec:model}}

The Hamiltonian of the $q$-state Potts model is
\begin{eqnarray}
  \mathcal{H} = -J\sum_{\langle i,j\rangle}\delta_{\sigma_i\sigma_j},
  \label{eq:hamiltonian}
\end{eqnarray}
where $\sigma = 1,2,\cdots,q$, $\langle i,j\rangle$ denotes nearest neighbors, $J>0$ (ferromagnetic), and $\delta$ is the Kronecker delta function. It is known~\cite{Wu1982} that this model on the square lattice goes through a phase transition at temperature 
\begin{equation}
T^{\rm exact}_c=\frac{J/k_B}{\ln(1+\sqrt{q})},
\label{eq:tcexact}
\end{equation}
and this transition is continuous for $q=2,3,4$ and first-order for $q>4$~\cite{Baxter1973}. In the marginal case $q=4$ there are logarithmic corrections to the power-law critical behaviors~\cite{Cardy1980}, and for larger $q$ the transition is weakly first-order for $q$ close to $4$. In Tab.~\ref{tab:corre_length}, the analytically calculated correlation lengths at the transition point~\cite{Klumper1989,Buffenoir1993} are listed for $5 \le q \le 10$. The five-state model has a very long correlation length, much larger than what can be achieved in MC simulations, and exhibits an extremely weakly first-order phase transition. Note that the correlation lengths in Tab.~\ref{tab:corre_length} are obtained assuming the invariance under the duality trasformation. The correlation lengths at the transition point approached from the disordered phase are likely different from those of the ordered phase, and the exact results in Tab.~\ref{tab:corre_length} would be between them.
%

\begin{table}[t]
\caption{\label{tab:corre_length}Correlation length at the transition point $\xi(q)$ according to Eq.~(4.46) in~\cite{Buffenoir1993}.}
\begin{ruledtabular}
  \begin{tabular}{|c||c|c|c|c|c|c|c|}
    $\ \ q\ \ $ & 4 & 5 & 6 & 7 & 8 & 9 & 10 \\ \hline
    $\ \ \xi\ \ $ & $\infty$ & 2512.2 & 158.9 & 48.1 & 23.9 & 14.9 & 10.6 \\
  \end{tabular}
\end{ruledtabular}
\end{table}

We perform MC simulation for the cases $q=5,6,7,8$ on periodic square lattices using the Swendsen-Wang algorithm~\cite{Swendsen1987}. The measurements are computed in 960 independent samples with $10^8$ times cluster updates after the convergence of the Markov process. The algorithm reduces the critical slowing down relative to single-spin Metropolis updates by flipping clusters whose size is comparable to the physical correlation length. Generally, we should not expect that the cluster algorithm is efficient close to a first-order transition point. We can understand this from the observation that the typical cluster size in one thermodynamic state may be different from that in the other one. If that is the case, when we are in the state with smaller cluster size, we need to wait until many clusters flip in some particular way so that the resulting configuration becomes a typical state of the phase with the larger correlation length. If the first-order transition in question is accompanied by spontaneous symmetry breaking, as in the Potts models, this condition becomes very severe as the system size increases, because the relevant correlation length to use in the above argument in the ordered phase is infinite. In other words, trying to change a typical configuration of a disordered phase with small correlation length into an ordered one by cluster flipping is similar to tossing many coins and hoping all of them land as heads.

By the above arguments, in simulations of systems at weakly first-order transitions, one may expect that the cluster algorithm should still be effective, as long as the system size is less than the correlation length and the nature of first-order transition is not yet so obvious. Nevertheless, it is known that even if the system size is much smaller than $\xi\approx 2500$ in $q=5$ simulations, the relaxation to the Boltzmann distribution is very slow~\cite{Peczak1989}. Although the reason for the slow MC dynamics is not understood, the $q$-state Potts model with $q\geq 5$ nevertheless exhibits such an extremely slow relaxation that it is practically impossible to simulate systems beyond $L=256$ (the largest size consider in our work here) even in the most weakly first-order case of $q=5$.

Turning to observables, we consider the order parameter defined as a complex magnetization,
\begin{eqnarray}
  m = \frac{1}{N}\sum_j e^{i\frac{2\pi}{q}\sigma_j},
  \label{eq:complex_magnetization}
\end{eqnarray}
and compute the expectation values of its square and fourth power, $\langle m^2\rangle=\langle |m|^2\rangle$ and $\langle m^4\rangle=\langle |m|^4\rangle$ respectively. Notice this order parameter can detect the $S_q$ symmetry breaking of Potts models without any problem though it is usually employed for $Z_q$ symmetry breaking. We can calculate these quantities using the standard method of improved (cluster) estimators in the Swendsen-Wang algorithm~\cite{Wolff1989}. We will analyze $\langle m^2\rangle$ as well as the fourth-order Binder ratio,
\begin{equation}\label{eq:binder4}
R_4 = \frac{\langle m^4\rangle}{\langle m^2\rangle^2},
\end{equation}
which at some continuous phase transitions approaches a step function as the system size is increased but has a divergent feature at the step if the transition is of first order.
%

\section{Detection through scaling exponents\label{sec:exponent}}

Here we first discuss the result of the FSS data-collapse approach, applied to a single large data set including many different system sizes, which plausibly gives a fictitious exponent and misleads us about the order of the phase transition. We then discuss an alternative FSS approach to study the flow of the effective (size dependent) exponents, a `curve-crossing method', where the dimensionless Binder ratio $R_4$ is considered for pairs of system sizes at the point where their $R_4$ values coincide close to the phase transition. We show the extrapolation in terms of the system size makes it possible to detect the first-order transition.

\subsection{Data-collapse analysis\label{subsec:collapse2}}

\begin{figure}
\includegraphics[width=8cm]{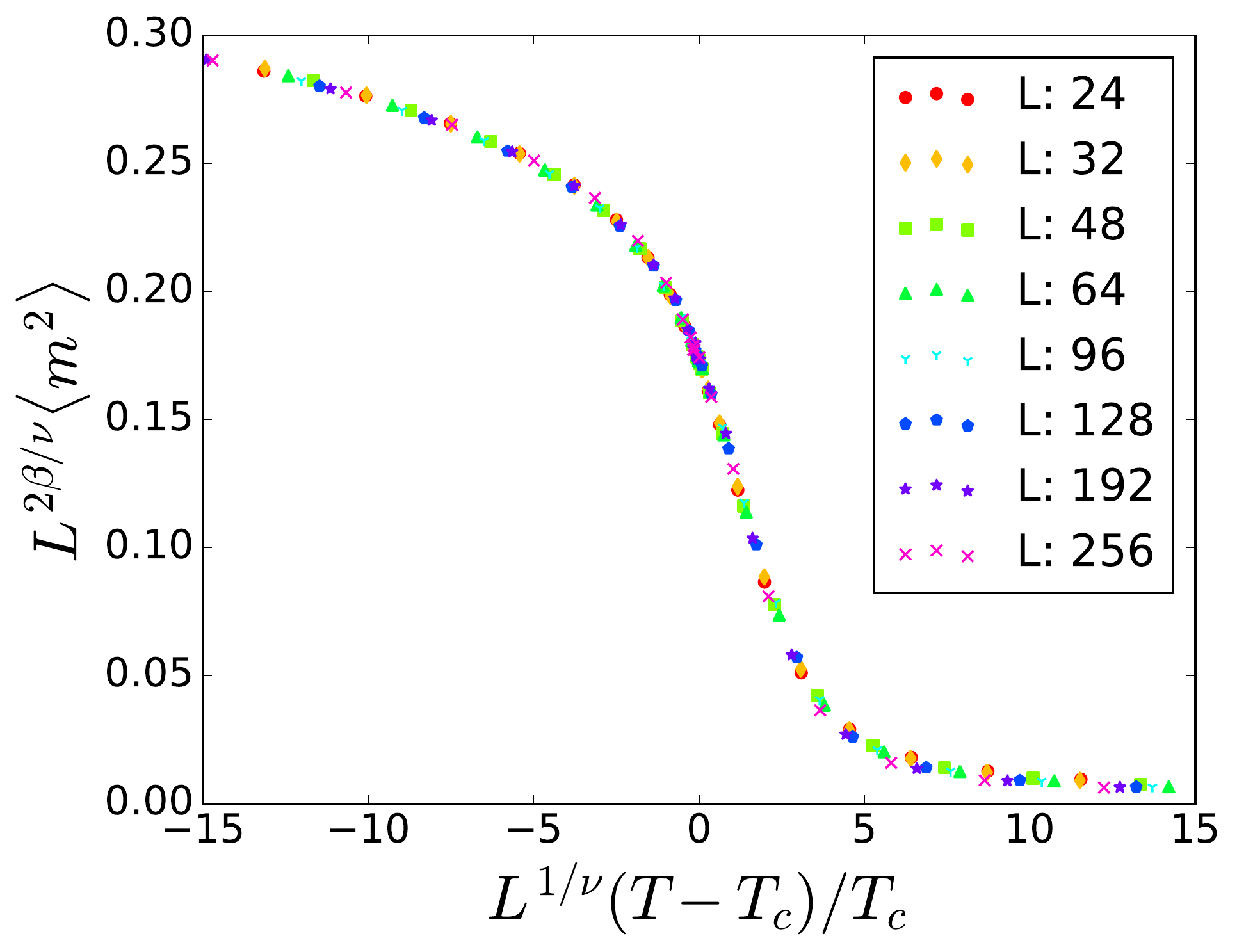}%
\caption{\label{fig:datacollapse}
Optimal FSS data collapse of the squared magnetization of the $q=5$ Potts model using a broad range of system sizes $L$. Note that the statistical error of each data point is smaller than the symbol size. The effective exponents obtained in this case are $\beta=0.070(8)$ and $\nu=0.61(5)$, and the transition temperature is $T_c=0.851(5)J/k_B$. The number of the parenthesis represents the statistical error estimated by BSA~\cite{Harada2011,Harada2015}, which are obviously too small to cover the correct values of exponents, $\beta = 0$ and $\nu = 0.5$.}
\end{figure}

First of all, Fig.~\ref{fig:datacollapse} we show the outcome for the $q=5$ model of an FSS data-collapse analysis of the squared order parameter, $\langle m^2 \rangle$, whose scaling dimension at a transition point is $2\beta/\nu$. Thus, we multiply $\langle m^2 \rangle$ by $L^a$, where $a$ is interpreted as an effective, adjustable value of $2\beta/\nu$, and plot the results against the scaled distance to the transition point, $L^{1/\nu}(T-T_c)/T_c$. Here $\nu$ and $T_c$, too, are treated as adjustable parameters. Then, using the estimated $a$ and $\nu$, we can evaluate $\beta$. We compare the results with the asymptotically (large-$L$) expected exponents, $\nu=1/d=1/2, \beta = 0$, and the value of $T_c$ for $q = 5$ from Eq.~(\ref{eq:tcexact}). To obtain the optimal effective exponents and $T_c$, i.e., to achieve the best collapse of data for a wide range of different system sizes, we use Bayesian Scaling Analysis (BSA)~\cite{Harada2011,Harada2015}. In Fig.~\ref{fig:datacollapse} we show an example with system sizes from $L=24$ to $256$. The data collapse here is so successful that the figure would seem to suggest, rather convincingly, a continuous phase transition with non-trivial values of the critical exponents, if we did not know the true first-order nature of the transition. It is known that a `pseudoscaling' like Fig.~\ref{fig:datacollapse} is also observed in various spin models which show weakly first-order transition (e.g., see Ref.~\onlinecite{Kamiya2010}). Though the exponents are clearly not correct, the value of $T_c$ agrees very well with the exact $q=5$ value $T_c \simeq 0.85153J/k_B$ from Eq.~(\ref{eq:tcexact}).
%

\subsection{Curve crossing method}

By gradually changing the system-size window used in the data-collapse analysis, as was done in Ref.~\onlinecite{Harada2013}, one can see the systematic trend and detect deviations, if any, from a conventional critical behavior. However, in this article we use an alternative method proposed by Fisher as "phenomenological renormalization"~\cite{Fisher1972,Luck1985}. In fact, the estimates of the exponents obtained by both method show essentially the same trend as a function of the system size, in terms of the existence of `cross-over' behavior we will explain later.

\begin{figure}
\includegraphics[width=7.5cm]{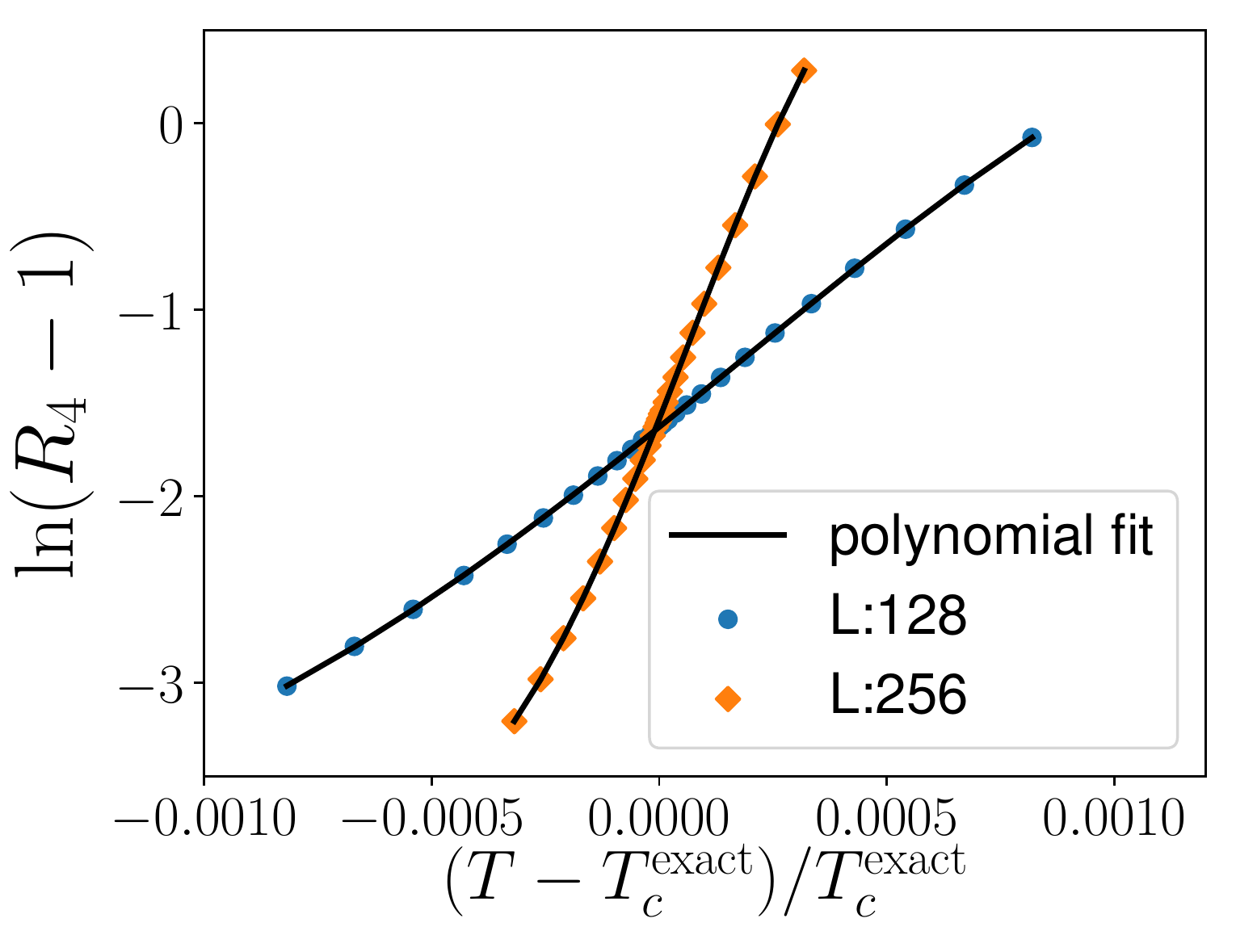}%
\caption{\label{fig:cross_example}
  The exponents $\nu$ and transition temperatures at the scale $L$ are computed as follows:
  (i) plot dimensionless quantities computed in the $L/2\times L/2$ and $L\times L$ lattices, generated randomly by Gaussian distribution using the averages and statistical errors of the MC results,
  (ii) and then an appropriate polynomial fitting makes it possible to calculate the crossing point of two curves and Eq.~(\ref{eq:crossing_point}); (iii) repeating (i) and (ii) and taking statistics, we can estimate the mean values of the exponents and $T_c$ with the statistical error.}
\end{figure}

In the phenomenological renormalization, one considers a dimensionless quantity, for which curves plotted versus the control parameter (here the temperature) for two different system sizes, $L_1$ and $L_2$ (for example of the form $L_1=L$ and $L_2=2L$ or $L_1=L$ and $L_2=L+\Delta L$ with a constant increment $\Delta L$), will cross each other at a point approaching the transition point as the system sizes are taken to infinity, and a corresponding critical value of the dimensionless quantity is approached in the vertical direction, see Fig.~\ref{fig:cross_example}. The flows of the horizontal and vertical crossing values, as well as the slopes at the crossing points, are governed by the exponent $\nu$ and corrections to scaling (the subleading exponent $\omega$). The method has been used extensively for systematic extrapolations of exponents in both classical and quantum statistical mechanics. The method was recently illustrated with detailed tests and applied to deconfined quantum-criticality in Ref.~\onlinecite{Shao2016}, and we here follow the same procedures to analyze the size flow of the effective exponent $\nu$ and $\beta$ for $q=5$.
%

For the dimensionless quantity, we consider the standard fourth-order Binder ratio of the order parameter, Eq.~(\ref{eq:binder4}). In practice, we evaluate $R_4$ on a dense grid of temperatures and use polynomial fits for interpolation to obtain the crossing points for system sizes $(L/2,L)$ as well as the slopes at the crossing point (from which $\nu$ is extracted). The polynomial fitting is better behaved with the quantity $\ln(R_4-1)$, which is linear in the vicinity of transition point, as we will show later in Sec.~\ref{sec:binder}. We then obtain the estimate of $\nu$ at the scale $L$ as
\begin{eqnarray}
  \frac{1}{\nu(L)} = \log_2 \left[ \cfrac{\cfrac{d}{dt}\ln\left[R_4(t,L)-1\right]}{\cfrac{d}{dt}\ln\left[R_4(t,L/2)-1\right]} \right]_{t=t_{\rm cross}},
  \label{eq:crossing_point}
\end{eqnarray}
where $t$ is the reduced temperature with $t_{\rm cross}$ corresponding to the crossing point of $\ln(R_4-1)$ at $L$ and $L/2$, which is computed from the fitted polynomials. The derivatives are also evaluated using the same polynomials. In addition, employing the $\nu$ and the crossing point, the exponent $\beta$ can also be computed as the ratio of slopes of squared magnetization at the crossing point of the Binder ratios:
\begin{eqnarray}
  \frac{1-2\beta(L)}{\nu(L)} = \log_2 \left[\cfrac{\cfrac{d}{dt}\langle m^2\rangle(t,L)}{\cfrac{d}{dt}\langle m^2\rangle(t,L/2)} \right]_{t=t_{\rm cross}}.
  \label{eq:crossing_point_beta}
\end{eqnarray}
The slope of $\langle m^2\rangle$ is also obtained through polynomial fitting of it.

\subsection{Analysis of size-dependent effective exponents}

\begin{figure}
\includegraphics[width=7.5cm]{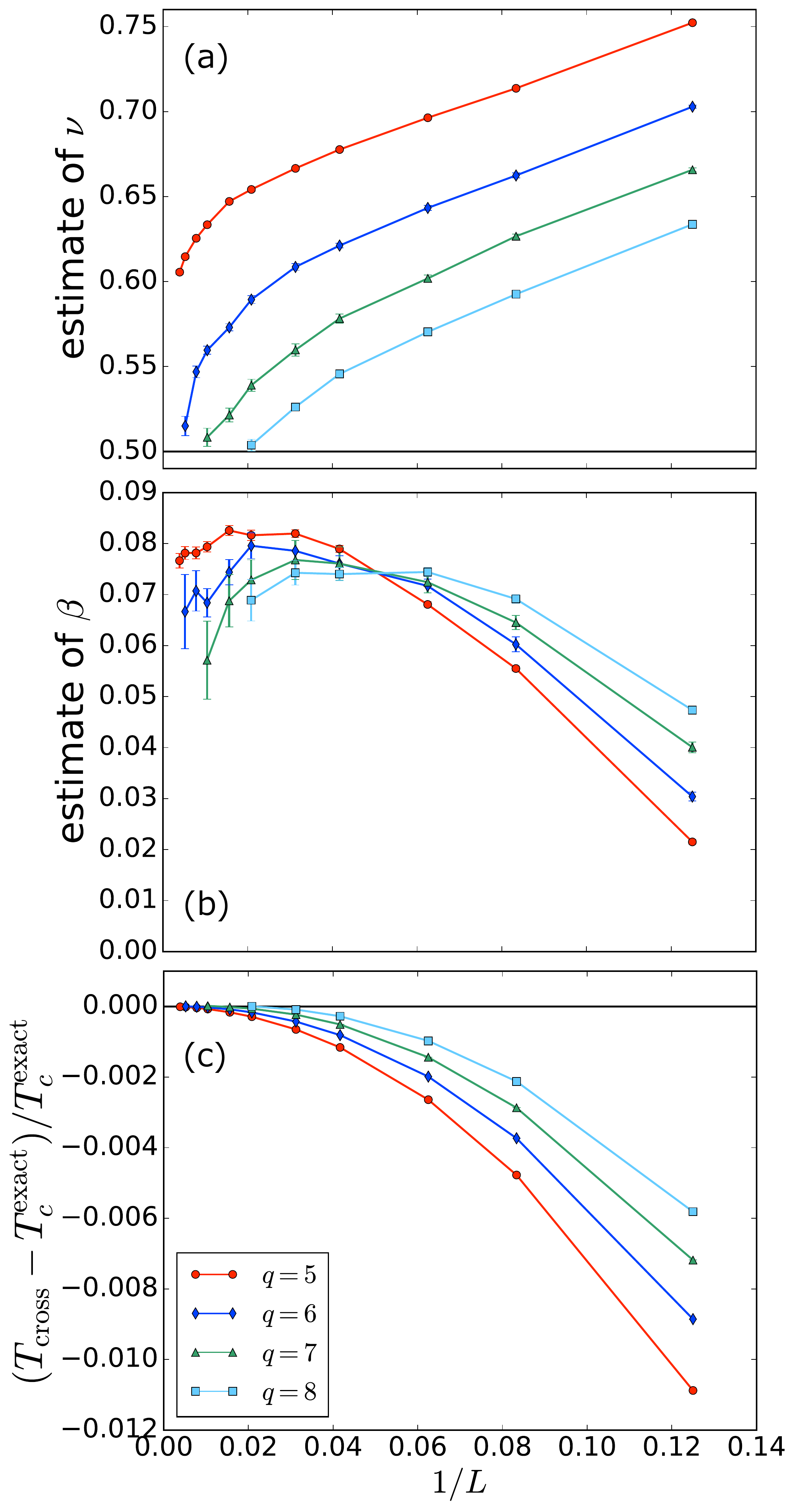}%
\caption{\label{fig:cross}
(a) Estimated exponents $\nu$ for $q=5,6,7,8$ obtained from crossing point analysis with system-size pairs $(L/2,L)$. The black horizontal line indicates the expected asymptotic value $1/d=1/2$.
(b) Estimated exponents $\beta$ for $q=5,6,7,8$ obtained from the Eq.~(\ref{eq:crossing_point_beta}).
(c) Deviation of the size dependent transition temperature $T_{\rm cross}$, defined as the Binder crossing point, and the exact transition temperature $T_c^{\rm exact}$ from Eq.~(\ref{eq:tcexact}).}
\end{figure}

In Fig.~\ref{fig:cross}, we show the size dependent estimates of the exponent $\nu$, $\beta$, and the transition temperature from the crossing point analysis. For $q = 6,7,8$ we can observe clearly in Fig.~\ref{fig:cross}(a) how the exponents approach the first-order value $\nu=1/2$, with a flattening-out to an eventual exponential size dependence expected for still larger system sizes. For $q=5$, as well, we observe a cross-over into what appears to be a similar rapid drop. On the other hand, the exponent $\beta$ shows clearly nonmonotonic behavior in Fig.~\ref{fig:cross}(b): upon increasing the system size, initially we see that the exponents deviate further away from the ultimately expected first-order values, $\beta=0$, but beyond some length scale, manifested as a maximum, they start to approach the correct values. Although the estimates at the largest $L$ are a little distant from the trivial value, an eventual drop to $0$ appears likely. In Sec.~\ref{sec:conclusions}, it will be pointed out that this nonmonotnicity may signal some interesting behavior in the renormalization group flow of the weakly first-order transitions.

An important question now is whether one can actually detect the first-order transition unambiguously by some kind of extrapolation of the size dependent quantities. To answer this question, let us pretend that the transition is continuous and see if any inconsistency results from that. In doing so, we consider the effect of the corrections to scaling that should generally exist if the transition is continuous. Now we focus on the case of the exponent $\nu$ in Fig.~\ref{fig:cross}. In the curve-crossing method, the asymptotic flow of $\nu$ toward the value in the relevant universality class has a finite-size correction of the form $\propto L^{-\omega}$, where $\omega>0$ is the exponent of the leading scaling-correction (irrelevant field)~\cite{Luck1985}. Extrapolations can then be carried out based on fits to this form, normally applied directly to the inverse value of the exponent obtained from the data according to from Eq.~(\ref{eq:crossing_point})~\cite{Shao2016}. As we have discussed above, in the first-order case the exponent ultimately approaches the value $\nu=1/d$ exponentially rapidly, but for a weakly first-order transition it may not be possible to reach the system sizes for which this behavior holds, e.g., in the case $q=5$ above. In the $L$ dependent curve-collapse method, one would also expect the same type of corrections if the procedure is carried out with systematically chosen groups of system sizes and the temperatures considered in the procedure are sufficiently close to $T_c$. With both methods, if sufficiently large systems are not accessible for the exponential convergence to be reached, one may still be able to fit at least some of the data to a conventional power-law correction, but the resulting exponent should then not be the correct one. 

\begin{figure}
\includegraphics[width=7.5cm,]{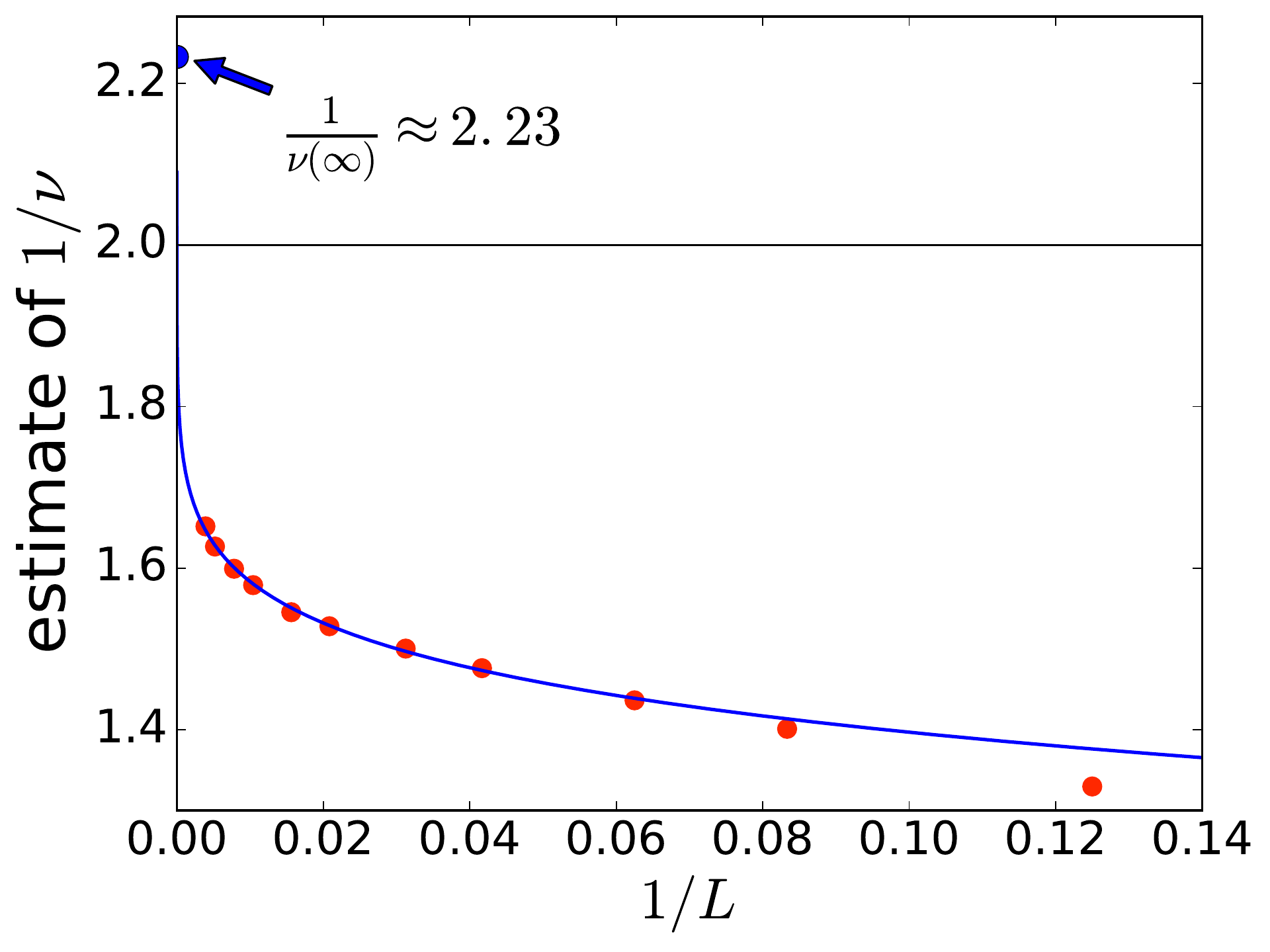}%
\caption{\label{fig:fitnu} 
Extrapolation of the inverse correlation-length exponent of the $q=5$ Potts model based on a power-law fit, $\nu^{-1}(L) = \nu^{-1}(\infty) + aL^{-\omega}$ (blue curve), to the data (red points) from Fig.~\ref{fig:cross}. The smaller sizes were excluded from the fit until a good $\chi^2$ value (the goodness of the fit defined in the standard way) was obtained; the minimum size is $L=16$ and the effective correction exponent of the fit is $\omega \approx 0.11$. Because the functional form used in the fit is ultimately wrong at a first-order transition, the extrapolated value $\nu^{-1}(\infty)\approx  2.23$ far exceeds the correct value $1/\nu=2$.}
\end{figure}

In Fig.~\ref{fig:fitnu} we show an example of such an extrapolation, using the curve-crossing results for $q=5$ from Fig.~\ref{fig:cross}. We find that a good fit to the power-law form can be obtained if the smallest system sizes are excluded, but with an anomalously small correction exponent, $\omega \approx 0.11$, and with an extrapolated value of $1/\nu$ that far exceeds the correct first-order value $2$. If more of the smaller system sizes are eliminated, the anomalous exponent values persist. These results suggest that the weakly first-order behavior can be detected based on this kind of anomalous behavior. Note that, one should never expect a value of $\nu$ less than $1/d$ at a continuous transition, and in some cases higher bounds can be obtained, e.g., from stability arguments in conformal field theories.
%

%

\section{Detection through Binder ratio\label{sec:binder}}

As we have seen above, strong indications of first-order behavior can be detected in the extrapolation of the effective critical exponents in terms of the system size. Another, well-known indicator of first-order transitions is the temperature dependence of the Binder ratio, Eq.~(\ref{eq:binder4}), in the neighborhood of the transition. In the thermodynamic limit, this quantity trivially approaches the value $1$ in the ordered phase, while in the disordered phase another value can be computed based on the Gaussian fluctuations of the order parameter (as is guaranteed in a system with a finite correlation length). Normally, for a finite-size system undergoing a continuous transition, the discontinuous jump between the constant values away from the transition point becomes a monotonic function which changes rapidly between the two values within a temperature window of size $\propto L^{-1/\nu}$. However, it is known that a specific kind of nonmonotonicity develops at a first-order phase transition~\cite{Binder1984}; phase coexistence within a temperature range scaling as $L^{-d}$ leads to a volume divergence, $R_4 \propto L^d$, in the vicinity of the step feature, on the disordered side of the transition.

While a nonmonotonic peak in the Binder ratio is often taken as a sign of a first-order transition, it should be stressed that also some continuous transitions are associated with such behavior. Examples in two dimensions include the Potts models with $q=3,4$, the Ashkin-Teller model, and the Ising model with both nearest and next-nearest-neighbor interactions~\cite{Jin2012}. Examples have also found in quantum magnets related to the phenomenon of deconfined quantum-criticality~\cite{Beach2009,Kaul2011}. However, in all these cases the peak is either not divergent or diverges very slowly with the system size, in some cases likely logarithmically~\cite{Jin2012}. Thus, to truly confirm that a transition is of first-order one should observe the volume divergence of $R_4$.

The behavior of Binder ratio for first-order transitions above referred to can be derived phenomenologically~\cite{Binder1981_Zphys,Vollmayr1993} by assuming a reasonable form of the probability distribution of an order parameter that breaks a $S_q$ symmetry in the ordered phase, as is the case with the $q$-state Potts model. In this subsection, we present a simpler derivation of this form of the Binder ratio. Also, the further insights helpful for detecting weakly first-order transitions is derived: the linearity of $\ln(R_4-1)$ in the scaled temperature.

\subsection{Phenomenological model of the Binder ratio}

We consider a first-order transition where the ordered phase has a $S_q$ symmetric order parameter in the form of a two-dimensional vector ${\boldsymbol m}=(m_x,m_y)$. The vector order parameter is equivalent to the complex magnetization defined in Eq.~(\ref{eq:complex_magnetization}). To take into account coexistence between the ordered and disordered phases and to be able to tune the system to either of the phases, we assume the following probability distribution of ${\boldsymbol m}$:
\begin{equation}
  P({\boldsymbol m}) \propto e^{tL^d}\exp\left({-\frac{{\boldsymbol m}^2}{2\sigma^2}}\right)
  + e^{-tL^d}\frac{1}{q}\sum_{p=1}^q \exp\left[ -\frac{({\boldsymbol m}-{\boldsymbol m}_p)^2}{2\sigma^2} \right]
  \label{eq:probability}
\end{equation}
where $t\propto T-T_c$ is the reduced temperature and the magnetization vector in the $p$-th ordered state is given by ${\boldsymbol m}_p = m_0\times(\cos(2\pi p/q),\sin(2\pi p/q))$. Gaussian fluctuations in a system with finite correlation length imply that the variance of the distribution Eq.~(\ref{eq:probability}) scales with the system size $L$ as $\sigma^2 = \chi_0 L^{-d}$. The form Eq.~(\ref{eq:probability}) of the distribution was previously discussed without invoking the size dependence~\cite{Sandvik2010}, and we here point out scaling behaviors when this aspect if included. The first term in Eq.~(\ref{eq:probability}) is dominant in the disorder phase ($t>0$) while the second term reflects the $S_q$ symmetry of the order parameter in a finite system when $t<0$. Note that $P({\boldsymbol m})$ is further constrained by the normalization $\int d{\boldsymbol m}P({\boldsymbol m})=1$. We also note that the ordered and disordered phases may, in principle, have different widths of the magnetization distribution, but we find that setting them equal to a common $\sigma$ given above does not significantly impact the conclusions we draw below.

Using Eq. (\ref{eq:probability}), we can compute the $n$th moment of the magnetization for even $n$;
\begin{equation}
  \langle m^n \rangle = \int \left\|{\boldsymbol m}\right\|^n P({\boldsymbol m}) d{\boldsymbol m}, 
\end{equation}
where the odd-$n$ moments vanish by symmetry. The Binder ratio can be expressed as
\begin{widetext}
\begin{eqnarray}
  R_4 &=& \frac{\langle m^4 \rangle}{\langle m^2 \rangle^2}
  = \left(1+e^{2tL^d}\right) \frac{8{\chi_0}^2\left(1+e^{2tL^d}\right)+8\chi_0L^d{m_0}^2+L^{2d}{m_0}^4}{\left[2\chi_0\left(1+e^{2tL^d}\right)+L^d{m_0}^2\right]^2},
\end{eqnarray}
\end{widetext}
which is independent of $q$. This Binder ratio has a cusp whose height and location are given by
\begin{eqnarray}
  {R_4}_{\rm peak} &=& \frac{\left( 8\chi_0+L^d{m_0}^2 \right)^2}{8\chi_0\left( 4\chi_0+L^d{m_0}^2 \right)},
  \label{eq:peak}
  \\
  t_{\rm peak} &=& \frac{1}{2}L^{-d}\ln\left( \frac{6\chi_0+L^d{m_0}^2}{2\chi_0} \right),
    \label{eq:peakpos}
\end{eqnarray}
respectively. The peak diverges asymptotically as $L^d$, and the scaled location $L^dt_{\rm peak}$ is slowly divergent as $\ln L$. The limits at high and low temperatures are given by
\begin{eqnarray}
  \lim_{t\rightarrow \infty}R_4 &=& 2\\
  \lim_{t\rightarrow -\infty}R_4 &=& \frac{8{\chi_0}^2+8{\chi_0}L^d{m_0}^2+L^{2d}{m_0}^4}{(2\chi_0+L^d{m_0}^2)^2}.
\end{eqnarray}
The low-temperature value can be consisitent when taking the limit $L\rightarrow\infty$: $\lim_{L\rightarrow\infty}\lim_{t\rightarrow -\infty}R_4=1$.

\begin{figure}[t]
\centering
\includegraphics[width=7.5cm]{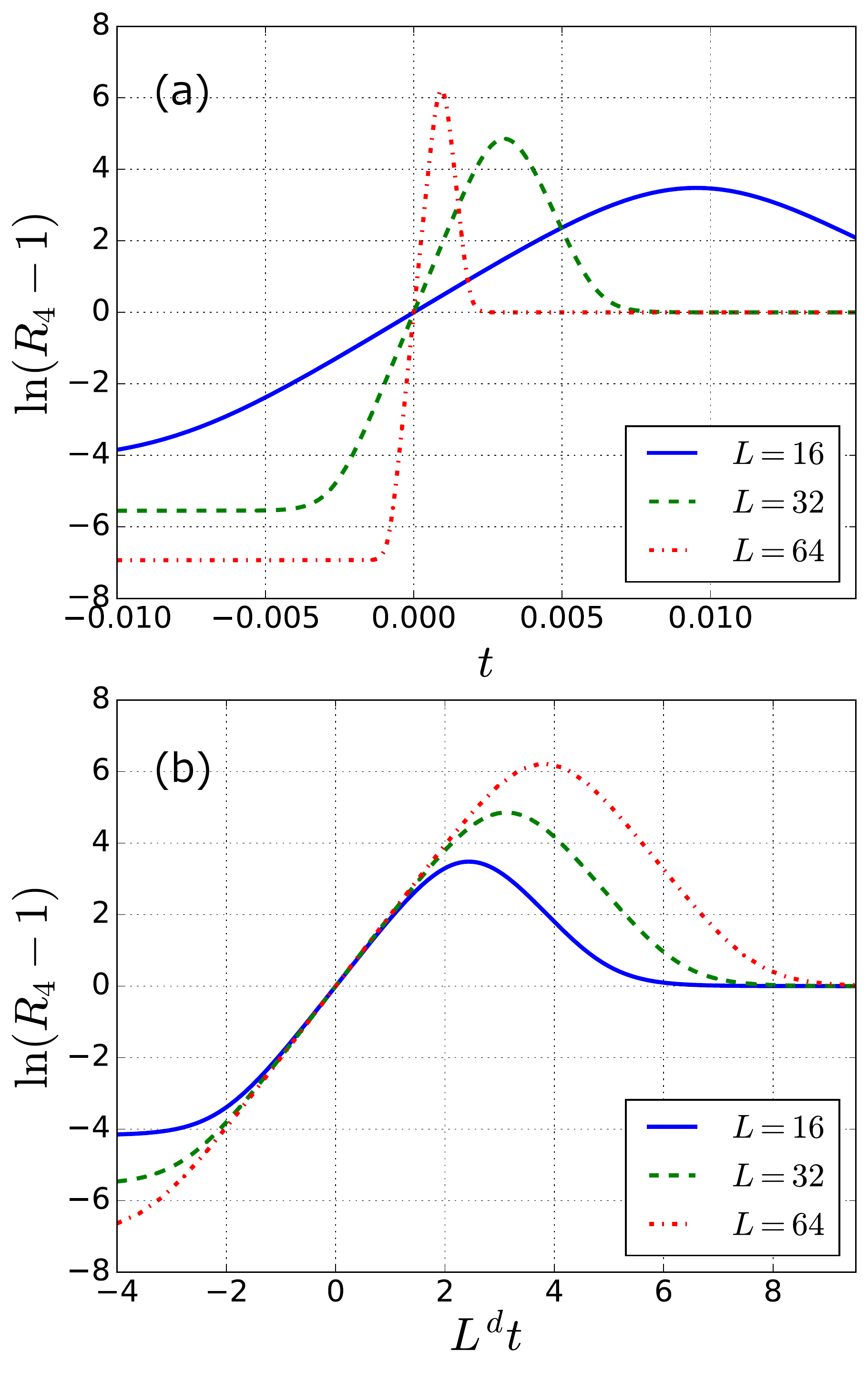}
\caption{\label{fig:R4lnR4} Binder ratios computed from the assumed probability distribution Eq.~(\ref{eq:probability}) with $d=2$, $m_0=1$ and $\chi_0=1$. In (a) results are shown for three different system sizes versus the reduced temperature, while in (b) the same results are shown versus the properly size-scaled temperature. We can observe the scaling of $\ln(R_4-1)$ is valid only in the linear regime.}
\end{figure}

We consider the behavior of $\ln(R_4-1)$,
\begin{widetext}
\begin{eqnarray}
  \ln\left(R_4-1 \right) = \ln
  \left\{
    \frac{4{\chi_0}^2\left(1+e^{2L^dt}\right)^2+4{\chi_0}\left(1+e^{2L^dt}\right)L^d{m_0}^2+e^{2L^dt}L^{2d}{m_0}^4}
         {\left[2{\chi_0}\left(1+e^{2L^dt}\right)+L^d{m_0}^2\right]^2}
         \right\},
\label{eq:fullformr4}
\end{eqnarray}
and expand it at $t=0$ as a function of $L^dt$;
\begin{eqnarray}
  \ln\left(R_4-1 \right) &=& \frac{2L^{2d}{m_0}^4}{\left(4\chi_0+L^d{m_0}^2\right)^2}L^dt
  -\frac{32\left({\chi_0}^2L^{2d}{m_0}^4\right)}{\left(4\chi_0+L^d{m_0}^2\right)^4}\left(L^dt\right)^2\nonumber\\
  &-&\frac{32\chi_0L^{2d}{m_0}^4\left(16{\chi_0}^3+40{\chi_0}^2L^d{m_0}^2+9\chi_0L^{2d}{m_0}^4+L^{3d}{m_0}^6\right)}{3\left(4\chi_0+L^d{m_0}^2\right)^6}\left(L^dt\right)^3 +O\left[\left(L^dt\right)^4\right].
  \label{eq:expansion}
\end{eqnarray}
\end{widetext}
From this expression we can conclude that the even-order terms are proportional to $L^{-2d}$ while the odd-order terms are proportional to $L^{-d}$, except the linear term. We show representative curves of the full form Eq.~(\ref{eq:fullformr4}) for three choices of $L$ in Fig.~\ref{fig:R4lnR4}.

One remarkable point indicated in the expansion Eq.~(\ref{eq:expansion}) is, that as a function of the scaled distance to the transiton point, $L^dt$, data collapse onto the same linear function in the vicinity of $L^dt=0$, but significant size dependence remains outside the linear regime, see Fig.~\ref{fig:R4lnR4}, as well as the volume-divergent peak. In other words, the FSS of $\ln(R_4-1)$ for first-order transition is valid only in the linear regime.
\subsection{Numerical results}

\begin{figure}[t]
\centering
\includegraphics[width=7.5cm]{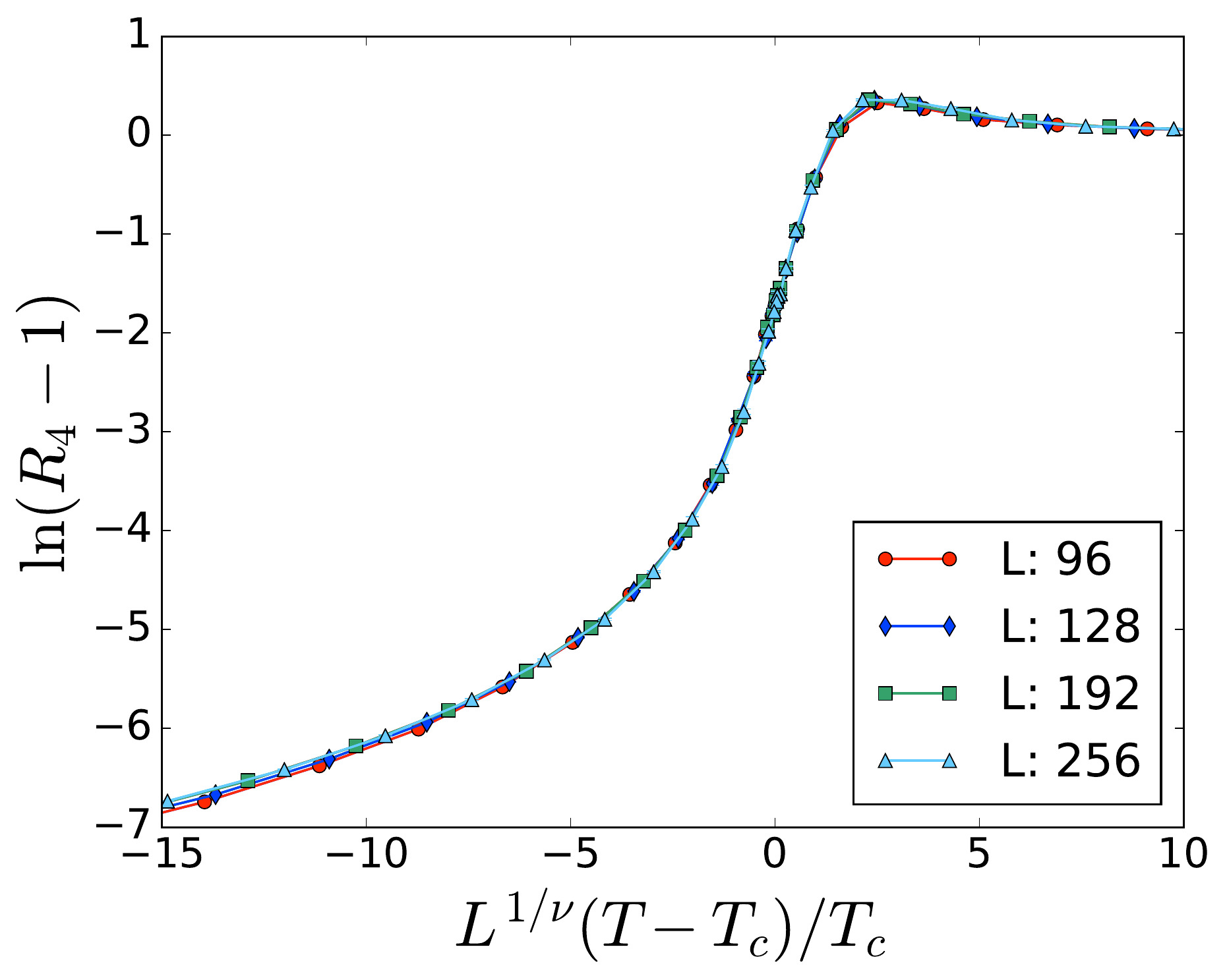}
\caption{\label{fig:q4} The data collapse of $\ln(R_4-1)$ for $q=4$ Potts model. The FSS is successful in the broad region including non-linear regime. The exponent and the transition temperature are $\nu=0.74(6)$ and $T_c=0.910(2)J/k_B$ respectively, which are computed by the BSA.}
\end{figure}

We first demonstrate the results of a continuous phase transition. The data collapse of $\ln(R_4-1)$ for $q=4$ Potts model obtained from MC simulation is shown in Fig. \ref{fig:q4}, whose exponent is computed by the BSA. We can clearly see that the broad range including non-linear regime falls on a single curve. Note that though it is known that $q=4$ Potts model has logarithmic correction to scaling~\cite{Cardy1980}, we assume the scaling without it to execute analyses as if we did not know any true nature of the phase transition, which is why we cannot obtain the true exponent $\nu=2/3$.
%

\begin{figure}
\includegraphics[width=7.5cm]{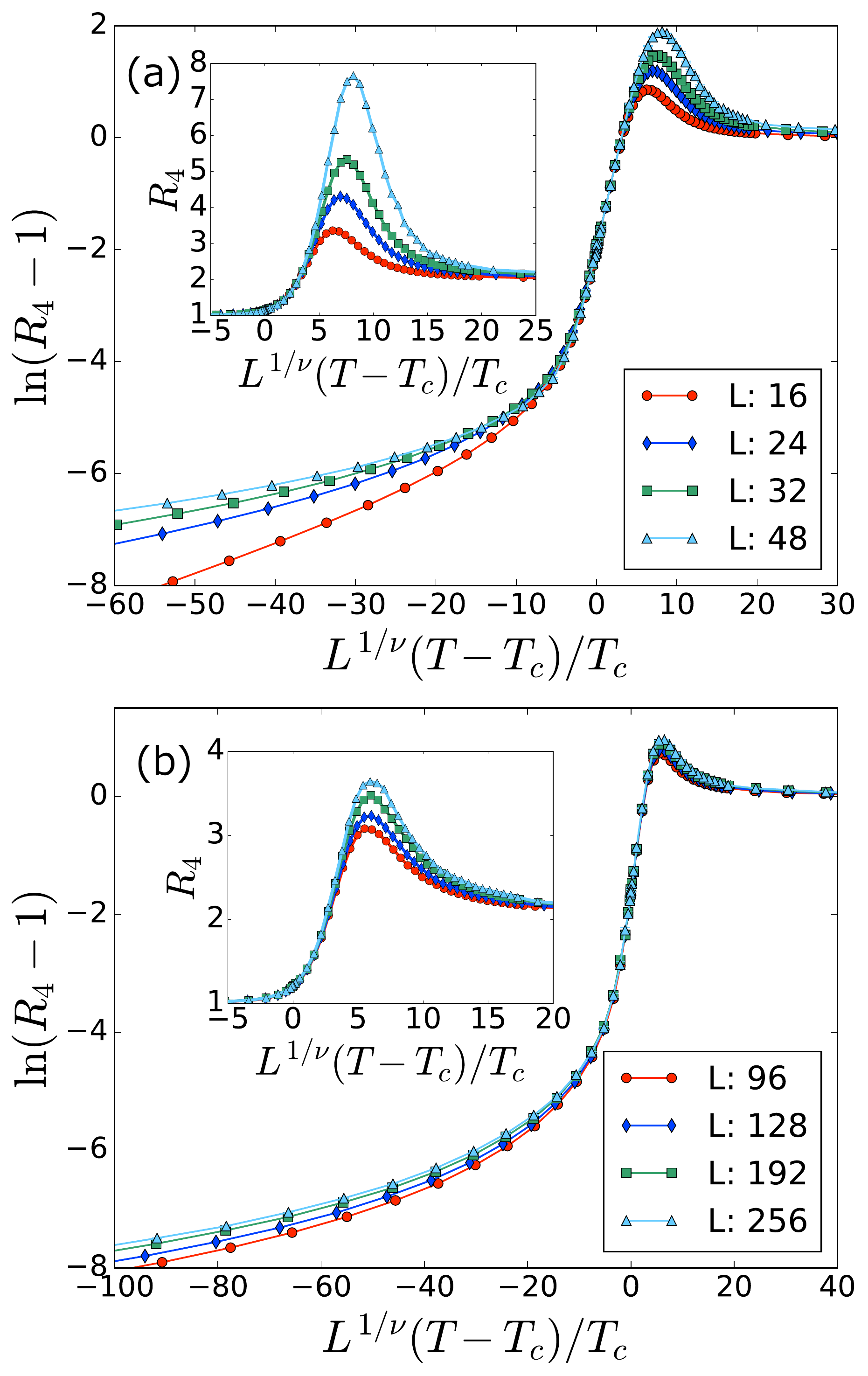}
\caption{\label{fig:lnR4}
The Binder ratio in the form of $\ln(R_4-1)$ of the (a) $q=8$ and (b) $q=5$ Potts models, plotted versus the scaled temperature for several system sizes. The results of the curve-crossing analysis (Fig. \ref{fig:cross} (a) and (c)) is used as the scaling exponents $\nu$ and transition temperatures $T_c$. The insets show the bare Binder ratio $R_4$ which is also scaled. The error bars are too small to be hiden in the symbols.}
\end{figure}

Figure \ref{fig:lnR4}(a) shows the Binder ratio of the eight-state Potts model. While the peak height grows with the system size, it is not quite proportional to the volume. This can again be understood as the system size not yet being sufficiently large compared to the correlation length. Though the growth is naturally even weaker for $q=5$, as shown in Fig. \ref{fig:lnR4}(b), we can still observe the peak sharpening for large $L$ (as seen more clearly in the inset of the figure). In addtion to the divergent peak height, the drift of peak position is qualitatively consistent with Eq.~(\ref{eq:peakpos}).

The most significant point in Fig.~\ref{fig:lnR4} is that the scaled $\ln(R_4-1)$ is linear in the vicinity of the origin and the data collapse is successful in this regime, while the points do not fall on the same curve in the non-linear region. These are clearly different from the case of $q=4$ shown in Fig. \ref{fig:q4} which does not show the size dependence outside the linear region. Therefore, in addition to the systematic size-dependence studies discussed in the above section, to examine whether the scaling region extends beyond the linear regime or not may be an alternative way of detecting the sign of first-order transition at much smaller system sizes than the correlation length.
%

\section{CONCLUSIONS AND DISCUSSION\label{sec:conclusions}}

By studying the two-dimensional $q$-state Potts models with $q \ge 5$ as known examples of systems with first-order transitions, we have shown that great care has to be taken when trying to determine whether a phase transition of unknown kind is continuous or weakly first order. In particular, for a very weak first-order transition, exemplified here by $q=5$, the standard data-collapse analysis of the order parameter can be deceptive, with beautiful scaling collapse obtaining for a wide range of system sizes with exponents quite far from the trivial exponents expected at all first-order transitions. However, signs of the first-order nature of the transition can still be seen, using the results on smaller lattices than the correlation length at the transition, if the exponents are analyzed carefully as a function of the system size. The behaviors normally expected at continuous transitions are violated, e.g., power-law extrapolations of the exponent $\nu$ deliver unphysical values. For these behaviors to be clearly manifested one still has to reach sufficiently large system sizes. If only far too small sizes are available, one may be easily misled by nonasymptotic effects and carry out erroneos extrapolations.

We also show that in first-order phase transitions the data collapse of $\ln(R_4-1)$ is successful only in its linear regime, which is guaranteed by phenomenological analysis. This behavior can be observed even in much smaller system sizes than the correlation length at the transition point. We demonstrate this method can diagnose correctly the order of the phase transitions of $q=4$ and $q=5$ Potts models.

We note that Ref.~\onlinecite{Lee1990} claimed that unambiguous signs of the first-order transition in the Potts model even for $q=5$ could be detected on lattices as small as $L=64$, by studying the double-peak (coexistence) structure in the free energy. However, similar to the divergent peak in the Binder ratio that we studied here, such features can also be observed in models with continuous transitions; an example is the Baxter-Wu model, which realizes the fixed point of the $q=4$ Potts model~\cite{Schreiber2005}. Since a reliable study is absent for the size-scaling of the peaks, the identification of the transition is not clear, and, like the order-parameter Binder ratio, the first-order forms require much larger sizes before they are seen unambiguously. Thus, we maintain that the method presented here is a more reliable method.

We would like to comment that the cross-over from the almost linear behaviors to the rapid drops of the exponent $\nu$ in Fig.~\ref{fig:cross}(a) and the nonmonotonic behavior of $\beta$ in Fig.~\ref{fig:cross}(b) may reflect pseudo-critical behavior in the renormalization group flow of weakly first-order transitions. One scenario for pseudo-critical behavior at a weakly first-order transition would be that the renormalization-group flow (which is captured in finite-size scaling) to the weakly first-order transition passes by the critical curve (which can be defined for continuous $q$ in the Potts models in the cluster representation of the partition function) ending at $q=4$. The proximity to the multicritical $q=4$ point might lead to effective exponents close to those at $q=4$ for a significant range of system sizes. The critical exponents of the $q=4$ Potts model are $\beta = 1/12\sim 0.083$. The values of $\beta$ observed in Fig.~\ref{fig:cross}(b) for $q=5$ are indeed not very far from the $q=4$ value, though there is no clear sign of convergence toward this value before the slow flow toward $\beta=0$ sets in. It should also be noted that there are logarithmic scaling corrections at $q=4$~\cite{Cardy1980}, and, therefore, it is not even easy to extract the exponents at that point~\cite{Jin2012}. Thus, it seems unlikely that one would actually ever be able to see any well-defined pseudo-critical scaling in the sense of almost $L$-independent exponents close to the $q=4$ values, even for $q=5$, as is also confirmed by our results.

Another possibility, recently advocated in Ref.~\onlinecite{Wang2017} in the context of deconfined quantum-criticality and with the Potts models presented as an example, is that the scaling for fixed $q$ close to $q=4$ may show pseudo-critical scaling due to the proximity of `nonunitary' critical points in the complex plane, which are known to exist~\cite{Cardy1980,Andrews1984} (see also Ref.~\cite{Blote2017} for a recent case where complex model parameters can change the renormalization flow in the vicinity of the $q=4$ Potts fixed point). In Ref.~\onlinecite{Gorbenko2018-1,Gorbenko2018-2}, the relation between weakly first-order phase transition and the complex fixed point in the imaginary axis is well reviewed and the comformal field theory for it is discussed. Here one might speculate that the almost linear behaviors of the exponent $\nu$ in Fig.~\ref{fig:cross}(a) for all the $q$-values studied may extrapolate to the $q$-dependent exponents of the corresponding nonunitary fixed points. However, although some properties are known of related nonunitary theories~\cite{Itzykson1986}, the exponents at the $q>4$ Potts points are not known, and we are therefore not in a position to test this intriguing scenario quantitatively. It would be interesting to more closely investigate the nonunitary fixed points and obtain reliable exponents for the values of $q$ studied here.
%

%

\begin{acknowledgments}
The authors thank H. Watanabe, Y. Motoyama, T. Obori, Y. Kato, and R. Kaneko for significant help and fruitful discussions, and also Cenke Xu, Wenan Guo, and T. Senthil for useful discussions. The numerical computations were performed on computers at the Supercomputer Center, ISSP, the University of Tokyo. S.I. is grateful to the support of Program for Leading Graduate Schools (ALPS). A.W.S. is supported by the NSF under Grant No.~DMR-1710170 and by the Simons Foundation. S.M. and N.K. are supported by MEXT as ``Exploratory Challenge on Post-K computer" (Frontiers of Basic Science: Challenging the Limits), and by ImPACT Program of Council for Science, Technology and Innovation (Cabinet Office, Government of Japan).
\end{acknowledgments}

\bibliography{bibliography}

\end{document}